\documentclass[12pt]{article}

\usepackage{scicite}

\usepackage{times}
\usepackage{graphicx}
\usepackage{float}

\usepackage{amsmath}
\usepackage{amssymb}

\usepackage{xcolor}

\newcommand\figeq[1]{\textcolor{black}{#1}}
\newcommand{\g}[1]{\mathbf{#1}}
\DeclareMathOperator{\Tr}{Tr} 

\newenvironment{sciabstract}{%
	\begin{quote} \bf}
	{\end{quote}}

\newcounter{lastnote}

\title{Single-Antenna Non-Line-of-Sight Matrix Imaging via  Reconfigurable Intelligent Surfaces}

\author
{Antton Goïcoechea$^{1\dagger\ast}$, François Sarrazin$^{1}$, Theodosios Karamanos$^{2,3}$, \\ Mathias Fink$^{4}$, Fabrice Lemoult$^{4}$, Matthieu Davy$^{1\ast}$\\
	\\
	\normalsize{$^{1}$Université de Rennes, CNRS, IETR, Rennes, France}\\
	\normalsize{$^{2}$Sorbonne Université, CNRS, Laboratoire GeePs, Paris, France}\\
	\normalsize{$^{3}$Université Paris-Saclay, CentraleSupélec, CNRS, Laboratoire GeePs, Gif-sur-Yvette, France}\\
	\normalsize{$^{4}$Institut Langevin, ESPCI Paris, Université PSL, CNRS, Paris, France}\\
	\\
    \normalsize{$\dagger$Present address: Institut Jean Le Rond d'Alembert, Sorbonne Université, CNRS, Paris, France}\\
	\normalsize{$\ast$Corresponding authors. E-mail: antton.goicoechea@dalembert.upmc.fr (A.G.)};\\ \ \normalsize{matthieu.davy@univ-rennes.fr (M.D.)}
}

\date{}

\begin{document} 

\baselineskip24pt

\maketitle 

\begin{sciabstract}
Modern imaging and sensing in complex environments, ranging from biomedical diagnostics to wireless communication, relies on accurately measuring and then controlling the wave propagation. Conventional approaches demand large arrays of antennas or transducers to reconstruct the full reflection or transmission matrix, enabling advanced protocols such as selective focusing or adaptive wave control. Yet, these arrays are expensive, bulky, and difficult to implement at microwave frequencies. Here, we show that a single transmitting–receiving antenna, when combined with a reconfigurable intelligent surface (RIS), can fully reconstruct the reflection matrix from far-field measurements, effectively transforming the RIS into a programmable synthetic antenna array. This approach allows high-fidelity imaging of complex scenes, selective focusing through clutter, and real-time tracking of moving targets. Our results establish RIS as a versatile, low-cost platform for matrix-based imaging, with broad implications for adaptive wave control, real-time sensing, and imaging in environments previously considered inaccessible.
\end{sciabstract}


\paragraph*{One-sentence summary}
{A single antenna combined with a reconfigurable intelligent surface enables high-fidelity imaging, selective focusing, and real-time tracking without costly antenna arrays.}

\section*{Introduction}
{Waves offer a powerful lens to probe and reconstruct complex environments, from microscopic structures to cluttered macroscopic scenes~\cite{ishimaruElectromagneticWavePropagation2017}. Wave-based imaging relies on analyzing the modifications that waves undergo when interacting with inhomogeneities, which can occur either in transmission, as in tomography or X-ray imaging, or in reflection, as in radar or ultrasonic backscattering. In the reflection scenario, for example, the goal is to reconstruct the reflectivity of a scene by measuring the time delays of scattered waves. This can naturally be achieved using arrays of transmitting and receiving probes. In its simplest form, an image is obtained by virtually back-propagating the recorded signals, either in the spectral or temporal domain~\cite{szaboDiagnosticUltrasoundImaging2004}. For monochromatic waves, the image resolution -- that is, the minimal resolvable pixel size -- inversely depends on the aperture of the transmitting and receiving arrays~\cite{goodmanIntroductionFourierOptics2003}, thus requiring large arrays with a high number of controllable elements for an efficient imaging technique.}

Beyond conventional imaging approaches, techniques based on the reflection matrix have emerged as powerful tools for non-invasive imaging in optics~\cite{kangImagingDeepScattering2015, badonSmartOpticalCoherence2016, yoonDeepOpticalImaging2020, Badon2020, yeminyGuidestarfreeImageguidedWavefront2021}, acoustics~\cite{varslotEigenfunctionAnalysisStochastic2004, lambertReflectionMatrixApproach2020, lambertDistortionMatrixApproach2020_2, bureauThreedimensionalUltrasoundMatrix2023}, and seismology~\cite{blondelMatrixApproachSeismic2018, giraudatMatrixImagingTool2024}. The reflection matrix gathers all reflection coefficients between a common set of transmitting and receiving channels, thereby encapsulating the full information content accessible to a given antenna array. By intelligently exploiting this information, one can correct for wavefront aberrations -- in a manner analogous to adaptive optics~\cite{babcockPossibilityCompensatingAstronomical1953, tysonPrinciplesAdaptativeOptics1991} -- or separate single -- from multiple-scattering contributions, among other advanced analyses. In addition to pure imaging applications, reflection-matrix measurements also provide a versatile framework for probing and characterizing complex scattering systems~\cite{tulinoRandomMatrixTheory2004}. However, while in acoustics and optics, large arrays composed of individually controllable elements -- typically hundreds in acoustics and millions in optics -- enable highly effective wavefront-shaping techniques~\cite{moskControllingWavesSpace2012, rotterLightFieldsComplex2017, caoShapingPropagationLight2022}, in the microwave regime, such multiple-input multiple-output (MIMO) architectures become prohibitively expensive and power-consuming as the number of controllable elements increases, owing to the inherent complexity of the associated hardware~\cite{bjornsonMassiveMIMOSystems2014, bjornsonMassiveMIMONonIdeal2015, gustavssonImpactHardwareImpairments2014}.

To overcome these limitations, one can rely on computational imaging techniques which provide efficient methods to reconstruct spatial information for low-cost and real-time imaging. These approaches exploit alternative encoding dimensions to capture spatial information indirectly. Pioneering demonstrations have shown that random masks can efficiently encode the necessary information of a scene. In optics, for instance, such masks can be generated by random spatial light patterns in single-pixel cameras~\cite{duarteSinglepixelImagingCompressive2008, edgarPrinciplesProspectsSinglepixel2019}.
The image is subsequently reconstructed by solving a linear inverse problem of the form $\g{y} = \g{H}\g{x}$, where $\g{y}$ is the $M \times 1$ measurement vector associated with the random masks, $\g{H}$ is a known $M \times N$ sensing (transmission) matrix, and $\g{x}$ is the unknown $N \times 1$ vector representing the reflectivity of the scene to be recovered. Crucially, compressive sensing theory even reveals that the number of measurements required to reconstruct a sparse scene can be drastically reduced since perfect recovery of a $K$-sparse scene -- that is, when only $K$ elements of $\g{x}$ are non-zero -- is theoretically achievable with only $\mathcal{O}(K \log(N/K))$ measurements. This demonstrates that computational and compressive imaging can together compensate for hardware limitations, enabling high-fidelity reconstructions from minimal data.

{This approach is particularly attractive in the microwave regime, where image reconstruction can be achieved using a single probe. This is reminiscent of the recent development of reconfigurable intelligent surfaces (RIS), which provide an energy-efficient means of generating tailored or random illumination patterns~\cite{kainaShapingComplexMicrowave2014, basarWirelessCommunicationsReconfigurable2019}. A RIS is a metasurface composed of electronically reconfigurable elements that enable precise control of wave reflection, in particular allowing the generation of complex or random wave fields.}
The desired sequence of structured patterns illuminating the scene can thus be synthesized by incorporating such RIS within a leaky cavity coupled to the probe. These dynamic metasurface antennas~\cite{yoo2018enhancing, shlezinger2021dynamic} enable unique possibilities for high-fidelity imaging of complex scenes~\cite{sleasmanImplementationCharacterizationTwoDimensional2021, imaniReviewMetasurfaceAntennas2020, sleasmanComputationalImagingDynamic2022}. However, this approach requires a lengthy calibration step, involving a near-field scan of the antenna for each mask, in order to determine the transfer matrix that links the metasurface configuration to the resulting scene illumination. Alternatively, the speckle patterns produced by a high-quality-factor (Q) leaky chaotic cavity, probed at different frequencies, can serve as random masks to encode the spatial content of the scene in the spectral domain. A large quality factor ensures that a large number of uncorrelated speckle patterns is generated within the operating bandwidth. In both cases, the scene is reconstructed by solving an inverse problem analogous to the single-pixel camera approach, enabling real-time imaging.

Yet another issue arises when trying to image objects outside of the detecting device's line of sight. In such a case, the idea is to use the scattered paths undergone by the wave instead of the direct path. Numerous techniques have been developed to enable imaging around corner, i.e. non-line-of-sight (NLOS) imaging~\cite{freundLookingWallsCorners1990,veltenRecoveringThreedimensionalShape2012,katzNoninvasiveSingleshotImaging2014, boger-lombardPassiveOpticalTimeofflight2019,faccioNonlineofsightImaging2020}, including with single-pixel cameras~\cite{musarraNonLineofSightThreeDimensionalImaging2019}. Despite recent advances in both NLOS and computational imaging, reconstructing the reflection matrix in the microwave regime -- which would enable advanced matrix-based imaging techniques -- remains elusive. For RIS-based antennas, one would ideally operate each element as if it were an individual antenna in an array. Yet, without a model describing the inter-element coupling, the full reflection matrix cannot be obtained. 

{In this article, we demonstrate that a single transmitting–receiving antenna coupled to a minimalist RIS with binary phase modulation can fully reconstruct the reflection matrix of a scene (Fig.~\ref{fig:principle}). We begin by modeling the complete scattering problem, representing the RIS as a system of coupled dipoles~\cite{faqiriPhysFadPhysicsBasedEndtoEnd2023, solExperimentallyRealizedPhysicalmodelbased2024, karamanosTopologyOptimizationMicrowave2024} to capture all inter-element interactions. In the absence of a scene, we show that recording the single-antenna reflection parameter for a series of random mask configurations, combined with a straightforward optimization, allows accurate retrieval of the full inter-element coupling. Once the RIS model is calibrated, the scene can be introduced, and—here comes the key result: the scene’s reflection matrix can be reconstructed analytically as if each RIS element were an independent emitter–receiver, without any additional optimization. With the reflection matrix in hand, one can immediately leverage the extensive literature on matrix-based imaging, for example assuming free-space propagation to reconstruct the spatial distribution of scatterers. Finally, we demonstrate imaging of a sparse scene containing only a few strongly reflective elements using a number of structured masks comparable to the number of RIS elements. Coupled with the high switching speed of the RIS, this approach opens a path toward real-time matrix-based imaging with a single transducer. While our experiments are performed in the microwave regime, the approach is fully general and applicable to all types of waves, including optics and acoustics.}

\section*{Results}
\subsubsection*{Theoretical model}
Our experimental setup, illustrated in \figeq{Fig.~\ref{fig:setup}A}, consists of a directive horn antenna illuminating two $8 \times 8$ 1-bit RIS, providing a total of $N_d = 128$ controllable elements. Each element can be electronically switched between two distinct states, ‘0’ and ‘1’, which induce a phase shift of approximately $\theta = \pi$ in the element’s reflection coefficient (see Materials and Methods). While the RIS are, in principle, capable of controlling both vertical and horizontal polarizations, only the vertical polarization is considered herein. All measurements are carried out at 4.1~GHz.

Random configurations of the RIS are generated by independently assigning each element to state ‘0’ or ‘1’. For each configuration, the {antenna's} reflection coefficient $\psi = S_{11}$ is measured at the transmitting–receiving antenna using a vector network analyzer. We consider that $\psi$ is made of three contributions: 1) a static reflection from the environment that does not depend on the RIS configuration $\Psi_0$; 2) the wave directly reflected by the RIS back toward the horn antenna; 3) the wave scattered by the RIS illuminating the scene to be imaged. After interacting with the target, the scattered field is reflected once more by the RIS and recorded by the antenna, as illustrated in \figeq{Fig.~\ref{fig:setup}B}.

To model the interaction of the incident wave with the RIS, we describe each element as a single dipole, such that the measured field can be expressed within the coupled dipoles framework~\cite{foldyMultipleScatteringWaves1945, laxMultipleScatteringWaves1951, DeVries1998}. Although this model constitutes a strong approximation of a real system composed of finite-size scatterers, it captures the essential physics required to accurately describe the field scattered by a RIS~\cite{faqiriPhysFadPhysicsBasedEndtoEnd2023, solExperimentallyRealizedPhysicalmodelbased2024, karamanosTopologyOptimizationMicrowave2024}. Beyond reproducing the measured transmission and reflection parameters for known RIS configurations, this approach also enables reliable predictions for new, untested RIS configurations.

Within the dipole model, the measured field for a given configuration $n$ can be expressed as the superposition of the three contributions as
\begin{equation}\label{eq: psi}
    \Psi^{(n)} = \Psi_0 + \g{G}^T \boldsymbol{\Gamma}^{(n)} \g{G}
    + \g{G}^T \boldsymbol{\Gamma}^{(n)} \g{R} \boldsymbol{\Gamma}^{(n)} \g{G},
\end{equation}
where $\g{G}$ is the $N_d \times 1$ vector of Green’s functions between the antenna and the RIS elements, $\boldsymbol{\Gamma}^{(n)}$ denotes the $N_d \times N_d$ effective reflectivity matrix of the RIS in configuration $n$, and $\g{R}$ is the reflection matrix of the scene expressed in the basis of the RIS elements. For the scalar case in three dimensions, the propagation of a wave in free space from point $i$ to point $j$, separated by the distance $r_{ij}$ is given by
\begin{equation}\label{G scalar}
    G_0(r_{ij}) = -\frac{e^{i k_0 r_{ij}}}{4 \pi r_{ij}}.
\end{equation}
Note that the full vector character of electromagnetic waves can be taken into account by using the dyadic Green's functions~\cite{jacksonClassicalElectrodynamics2009}. Since in our experiments the environment is rather uncluttered, we expect 
{each element of $\g{G}$ to be equal to its corresponding freespace $G_0(r)$ value.}

The term $\g{G}^T \boldsymbol{\Gamma}^{(n)} \g{G}$ in \figeq{Eq.~(\ref{eq: psi})} represents the wave scattered by the RIS directly back toward the antenna, without interacting with the scene. The matrix $\boldsymbol{\Gamma}^{(n)}$ fully accounts for the coupling between RIS elements and can be expressed as~\cite{solExperimentallyRealizedPhysicalmodelbased2024, karamanosTopologyOptimizationMicrowave2024}
\begin{equation}\label{eq: refl ris}
    \boldsymbol{\Gamma}^{(n)} = \boldsymbol{\alpha}^{(n)}\left[\mathbf{1} - \g{G}_{dd}\boldsymbol{\alpha}^{(n)}\right]^{-1}.
\end{equation}
The $N_d \times N_d$ matrix $\g{G}_{dd}$ describes all mutual interactions between elements, encompassing both near-field and far-field coupling, as well as contributions mediated by the substrate or by scattering from the surrounding environment. The diagonal matrix $\boldsymbol{\alpha}^{(n)}$ contains the individual polarizabilities of the RIS elements for configuration $n$. Since each element can be switched between two states, $0$ and $1$, they can be characterized by either $\alpha_0$ or $\alpha_1 = \alpha_0 r e^{i\theta}$. In the following, we assume that all elements are identical so that $\alpha_0$ and $\alpha_1$ do not depend on the element.

Finally, the third term in \figeq{Eq.~(\ref{eq: psi})} accounts for the interaction with the scene, encapsulated in the reflection matrix $\g{R}$ which is expressed in turn in the basis of the RIS elements. Our objective is to experimentally retrieve $\g{R}$ in order to reconstruct an image of the scene under the assumption of free space propagation between the RIS and the scene. 
Thus, we mathematically consider $\g{R} = \g{G}_0^T \boldsymbol{\gamma} \g{G}_0$, with $\boldsymbol{\gamma}$ the reflectivity of the scene. A standard confocal image~\cite{montaldoCoherentPlanewaveCompounding2009} is then obtained by applying free space propagators on each side: $\mathcal{I} = |\g{G}_0^* \g{R} \g{G}_0^{\dagger}|^2 \simeq |\boldsymbol{\gamma}|^2$.

\subsubsection*{Calibration}
{In the most general scenario, even in the absence of a scene to be imaged,} both the vector of Green's functions $\g{G}$ and the parameters of the RIS (namely $\alpha_0$, $\alpha_1$ and $\g{G}_{dd}$) are unknown. Considering that the coupling matrix $\g{G}_{dd}$ is symmetric, the number of unknown complex parameters is $N_{\mathrm{un}} = N_d + N_d(N_d+1)/2 + 2$. We therefore begin with a calibration procedure performed in the absence of any target, {\it i.e.} without the last term in \figeq{Eq.~(\ref{eq: psi})}, to determine these quantities. This calibration relies on a gradient-descent optimization process which is fully described in the Supplementary Material. The loss function minimized is the mean absolute error 
\begin{equation}
    \mathcal{L}(\boldsymbol{\Psi}_c,\hat{\boldsymbol{\Psi}}_c) = \left\langle \left| \boldsymbol{\Psi}_c - \hat{\boldsymbol{\Psi}}_c \right| \right\rangle_N
\end{equation}
where $\langle \cdots \rangle_N$ denotes averaging over $N$ random RIS configurations, $\boldsymbol{\Psi}_c$ is the vector of measured fields and $\hat{\Psi}_c$ is the field estimated from the coupled dipole approximation: $\hat{\Psi}_c^{(n)} = \Psi_0 + \g{G}^T \boldsymbol{\Gamma}^{(n)} \g{G}$. The loss function is computed from $N$ random RIS configurations, with $N > N_{\mathrm{un}}$. In this work, the retrieval of the parameters is performed using the Adam optimization scheme~\cite{kingmaAdamMethodStochastic2014}.

In such an optimization, the initialization of the parameters is crucial. We arbitrarily set $\alpha_0 = 1/k_0$, where $k_0$ is the free-space wavenumber. Multiplying $\alpha_0$ by a complex coefficient would simply renormalize $\g{G}_{dd}$, so this choice does not affect the model. Moreover, as the RIS elements are designed to induce a $\theta=\pi$ phase shift between their two states, we initialize $\alpha_1 = -\alpha_0$. As we will see later in the optimization result, this assumption is almost verified for our RIS. 

We then approximate the following terms using the first-orders Taylor expansion of $\boldsymbol{\Gamma}^{(n)}$. In this framework, the static contribution is the ensemble-averaged field, $\Psi_0 = \langle \Psi_c \rangle_N$. The first-order neglects inter-element coupling ($\g{G}_{dd}$) so that the effective reflectivity simplifies to $\boldsymbol{\Gamma}^{(n)} \simeq \boldsymbol{\alpha}^{(n)}$, leading to $\hat{\Psi}_c^{(n)} = \Psi_0 + \g{G}^T \boldsymbol{\alpha}^{(n)} \g{G}$. Since $\boldsymbol{\alpha}^{(n)}$ is diagonal, this problem is linear in $\g{G}^2 = \g{G}^T \g{G}$ and can be solved by inverting an $N \times N_d$ matrix given by the diagonal elements of $\boldsymbol{\alpha}^{(n)}$ for each configuration (see Supplementary Material). However, because $\hat{\Psi}_c^{(n)}$ depends on $\g{G}^2$ rather than $\g{G}$, the retrieved elements of $\g{G}$ exhibit a phase ambiguity of $\pi$. We lift this ambiguity by assuming phase continuity across the RIS and thereby correcting the $\pi$ phase shifts. Finally, we estimate the coupling matrix $\g{G}_{dd}$ via the second-order term of Taylor expansion $\boldsymbol{\Gamma}^{(n)} \simeq \boldsymbol{\alpha}^{(n)} + \boldsymbol{\alpha}^{(n)} \g{G}_{dd} \boldsymbol{\alpha}^{(n)}$ using a second matrix inversion (see Supplementary Material). We emphasize that this initialization requires only the solution of linear inverse problems, which is achieved through matrix pseudo-inversions. Its effectiveness stems from the RIS operating in free space. In a more complex environment, higher-order terms in the Taylor expansion of $\boldsymbol{\Gamma}^{(n)}$ would become more significant because of important inter-element couplings, making this initialization less robust.

{Out of the $2.5 \times 10^4$ random configurations that were measured, only $N=2^{14}=16,384$ were used for the gradient-descent optimization. The parameters obtained from this optimization are shown in Fig.~\ref{fig:calibration}.} {Note that we use a high number of total configurations in order to compare our optimized model with a large number of `unseen' configurations.}
We observe that, despite performing the measurements outside an anechoic environment, the Green’s functions between the antenna and the RIS, shown in \figeq{Fig.~\ref{fig:calibration}A and B}, is close to free-space propagation (\figeq{Eq.~(\ref{G scalar})}). Quantitatively, the correlation coefficient between the optimized Green’s functions and their free-space counterparts reaches $\sim 0.995$. The retrieved contrast between the two element states $\alpha_1 = r\alpha_0 e^{i\theta}$ (on) and $\alpha_0$ (off) is also in excellent agreement with expectations, with $r = 0.92$ and $\theta = \pi$. Furthermore, the estimated coupling matrix $\g{G}_{dd}$ (\figeq{Fig.~\ref{fig:calibration}C}) exhibits the anticipated behavior, with stronger interactions between neighboring elements.
Interestingly, these couplings are predominantly radiative, as their magnitude scales inversely with distance (\figeq{Fig.~\ref{fig:calibration}D}). Although matrix imaging is the ultimate objective, we emphasize that this approach also provides a powerful means to characterize a RIS comprehensively within a single experiment. Remarkably, the calibrated model can now predict the measured field for RIS configurations not included in the optimization process, achieving a mean absolute error as low as $0.2\%$ (see \figeq{Fig.~\ref{fig:calibration}E}). A small initial error is crucial for an accurate reconstruction of the reflection matrix in the following. Finally, we back-propagate the Green's vector $\g{G}$ assuming free space in \figeq{Fig.~\ref{fig:calibration}F}, which shows the location of the antenna.

\subsubsection*{Measurement of a reflection matrix}
We now turn to the estimation of the reflection matrix $\g{R}$ {in the presence of a scene that we want to image}. To isolate the contribution of the scene, we perform differential measurements defined as $\Delta \Psi^{(n)} = \Psi^{(n)}  - \Psi_c^{(n)} $, where $\Psi^{(n)} $ and $\Psi_c^{(n)}$ denote the measured fields in the presence and absence of the target, respectively. We note here that since the calibration allows for the estimation of unseen configurations with good precision, it is not necessary to perform this step with the same RIS configurations as before. From \figeq{Eq.~(\ref{eq: psi})}, this differential field can be expressed as $\Delta \Psi^{(n)} = \g{G}^T \boldsymbol{\Gamma}^{(n)} \g{R} \boldsymbol{\Gamma}^{(n)} \g{G}$. Since this relation is linear in $\g{R}$, we recast it as
\begin{equation}
    \boldsymbol{\Delta \Psi} = \g{H} \, \mathrm{vec}\left( \g{R} \right)
\end{equation}
where $\g{H}$ is an $N \times N_d^2$ matrix linearly relating the vector of differential measurements $\boldsymbol{\Delta \Psi}$ to the vectorized form of the reflection matrix $\mathrm{vec}(\g{R})$.  
The reflection matrix is then retrieved via the pseudo-inverse of $\g{H}$, denoted $\g{H}^+$, such that $\mathrm{vec}(\g{R}) = \g{H}^+ \boldsymbol{\Delta \Psi}$. Eventually, $\g{H}^+$ is computed using the singular value decomposition (SVD) with appropriate regularization to ensure numerical stability (see Supplementary Material for details). Note that in practice, we leverage the symmetry of $\g{R}$ to estimate only $N_d(N_d+1)/2$ terms and therefore accelerate the matrix pseudo-inversion.

\subsubsection*{Imaging of two scatterers}
Next, the reconstruction of the reflection matrix is applied to a scene consisting of two metallic plates of dimensions $5 \times 5~\mathrm{cm^2}$ placed in front of the RIS. The $\g{R}$-matrix is reconstructed from $N = 2.5 \times 10^4$ random configurations. We intentionally choose $N$ much larger than the number of independent elements of $\g{R}$ to be estimated $N_d(N_d+1)/2 = 8256$, ensuring that the inverse problem is overdetermined and that the matrix $\g{H}$ is full rank.

In \figeq{Fig.~\ref{fig:2targets}A}, we present the result of the optimization process, showing excellent agreement between the estimated and measured fields, with a mean absolute error of only $1.2\%$. The structure of the reconstructed reflection matrix $\g{R}$ is displayed in \figeq{Fig.~\ref{fig:2targets}B and C}. It exhibits a clear block structure composed of four large $64\times 64$ square regions. The diagonal blocks correspond to reflections involving the same RIS -- representing signals that are both transmitted and received by the same RIS -- while the off-diagonal blocks correspond to cross-interactions between the two RIS. Each large block is further subdivided into $8\times 8$ square sub-blocks, which can be interpreted analogously, but this time at the level of individual RIS elements. In the single-scattering regime, the sub-blocks of $\g{R}$ are expected to exhibit coherence along their antidiagonals~\cite{aubryRandomMatrixTheory2009}; however, given that each synthetic array comprises only eight elements per axis, this feature is less pronounced in the present configuration.

We then perform a SVD of the reflection matrix, $\g{R} = \g{U}\boldsymbol{\Sigma}\g{V}^{\dagger}$ (\figeq{Fig.~\ref{fig:2targets}D}). In the single-scattering regime, each subwavelength scatterer is associated with a single dominant singular value $\sigma_i$, whose corresponding singular vector $\g{U}_i$ represents the Green’s function linking the transmitting probes (here, the RIS elements) to that specific scatterer~\cite{Prada1994}. The SVD thus enables the extraction of individual wavefronts that selectively focus on each target. By contrast, an extended scatterer gives rise to multiple non-zero singular values, reflecting its distributed spatial extent.

In our experiment, as expected, two dominant singular values clearly emerge, corresponding to the two metallic plates. The remaining singular values, however, do not vanish entirely. We attribute this residual contribution to imperfections in the reconstruction of $\g{R}$ as well as to multiple scattering paths between the RIS and the targets that are not captured within our single-scattering formalism. The phase masks associated with the first two singular vectors, $\g{U}_1$ and $\g{U}_2$, shown in \figeq{Fig.~\ref{fig:2targets}E and F}, exhibit distinct spatial patterns. Each one exhibits a distinct spatial pattern, consistent with focusing at two different target locations.

For imaging purposes, we assume free-space propagation between the RIS and the targets. A confocal image $\mathcal{I}(r)$ is then reconstructed by back-propagating the reflection matrix $\g{R}$. This operation relies on the vector of Green’s functions $\g{G}_0(r)$, which describes the propagation between each RIS element and the point $r$ in the image plane. The image is obtained as $\mathcal{I}(r) = \left|\g{G}_0^\dagger(r)\, \g{R}\, \g{G}_0^*(r)\right|$.

The confocal image shown in \figeq{Fig.~\ref{fig:2targets}G} clearly exhibits two bright spots, as expected. We can further verify that the first singular vectors indeed correspond to each target by back-propagating them. The image obtained by back-propagating the vector $\g{U}_i$ is computed as $\mathcal{I}(r) = \left|\g{G}_0^\dagger(r)\, \g{U}_i\right|^2$. Each resulting image reveals a single bright spot at the position of the corresponding target (see \figeq{Fig.~\ref{fig:2targets}H and I}). This demonstrates our ability to selectively image individual targets, a key advantage of retrieving the full reflection matrix rather than a single image. Moreover, since $\g{U}_1$ and $\g{U}_2$ provide the optimal phase laws for focusing on each target, one can selectively illuminate or focus on them by appropriately configuring the RIS~\cite{karamanosTopologyOptimizationMicrowave2024}, {which can have applications in the context of optimizing the wireless communication channels}.

\subsubsection*{Imaging complex targets}
For targets whose dimensions are not small compared to the wavelength, the SVD no longer reveals a single dominant singular value but rather several, the number of which is related to the target’s size~\cite{toraldodifranciaDegreesFreedomImage1969, goriShannonNumberDegrees1973}. To a first approximation, the number of non-zero singular values corresponds to the number of resolution cells contained within the target. We demonstrate the performance of our method on more complex shapes by imaging each of the letters forming the word “IETR,” as shown in \figeq{Fig.~\ref{fig:IETR}}. For each letter, the measured field is accurately predicted, with a mean absolute error of approximately $2\%$, as illustrated in \figeq{Fig.~\ref{fig:IETR}A} for the letter ``R".

The structure of the reflection matrix shown in \figeq{Fig.~\ref{fig:IETR}B and C} for the letter ``R" resembles the one obtained for the two small targets case (\figeq{Fig.~\ref{fig:2targets}B and C}) although one can now see more complex features due to the complexity of the scene itself. We readily observe that the singular value spectrum of $\g{R}$ depends on the target, as shown in \figeq{Fig.~5D}. The number of dominant singular values correlates strongly with the spatial complexity of each letter. In all cases, we are able to reconstruct clear images of the letters (\figeq{Fig.~\ref{fig:IETR}E}), demonstrating the robustness of the method even for intricate shapes. Beyond conventional imaging, the extraction of the singular vectors also enables selective focusing of the wave field onto the brightest regions of the object.

The analysis of the mean absolute error of the estimated field, provided in the Supplementary Material, shows that a faithful reconstruction is achieved when the number of random configurations $N$ is larger than the number of independent elements of the $\g{R}$-matrix, $N > N_d(N_d+1)/2$, ensuring a full-rank $\g{H}$-matrix and an overdetermined problem.

\subsubsection*{Following a moving target}
The large number of configurations previously used to reconstruct the full reflection matrix is hardly compatible with real-time imaging. However, for a sparse scene containing a single dominant scatterer, it is sufficient to accurately reconstruct only the first singular vector, $\g{U}_1$, instead of the entire matrix $\g{R}$. The number of unknown coefficients to be retrieved is therefore reduced from $N(N+1)$ (for $\g{R}$) to only $N$ (for $\g{U}_1$). This simplification is valid when the first singular value of $\g{R}$ clearly dominates the others, providing an accurate description of the target position.

To enhance the signal-to-noise ratio, we further select the configurations yielding the maximal transmission to the imaging area. By focusing the signal on the imaging area, we expect a better signal to noise ratio as well as a more accurate description of the experiment with our physical model. We compute for each configuration the amplitude of the transmission to the focal plane as 
$T^{(n)} = \int_r \mathrm{d}r \, \left| \g{G}_{0}^T \boldsymbol{\Gamma}^{(n)} \g{G}_{\mathrm{im}} \right|$,
where $\g{G}_{\mathrm{im}}$ denotes the vector of Green’s functions between the RIS elements and the image pixel at position $r$ (see Supplementary Material). Out of $1.5\times10^4$ total configurations,
we select the $N$ configurations that maximize $T^n$. Importantly, this pre-selection step is performed after the calibration one but prior to measurements with the target. Thus, it depends only on the imaging domain, and not on the actual target position.

In \figeq{Fig.~\ref{fig:Moving}A}, we show the correlation coefficient $\mathcal{C} = |\g{U}_{1} \cdot \g{U}_{1,\mathrm{ref}}|$ as a function of the number of random configurations $N$, where $\g{U}_1$ is estimated from $N_c$ configurations and $\g{U}_{1,\mathrm{ref}}$ is a reference obtained from $1.5\times10^4$ configurations. Remarkably, a correlation as high as $C = 0.5$ is reached with only $N = 2N_d = 128$ selected configurations, compared to $C = 0.28$ for unselected random configurations. For these selected configurations, the first singular vector clearly dominates the second one (see \figeq{Fig.~\ref{fig:Moving}B}). While random configurations produce a noisy image that fails to localize the scatterer (\figeq{Fig.~\ref{fig:Moving}C}), the scatterer position is clearly retrieved for selected configurations, as shown in \figeq{Fig.~\ref{fig:Moving}D–E}, for different target positions. {We finally perform an experiment in which the target is displaced in the $xy$-plane at each time step. We use $N=1000$ pre-selected configurations to form the image, which yields a trajectory that follows closely the one imposed on the target, as shown in  \figeq{Fig.~\ref{fig:Moving}F}.}

Our SVD-based approach therefore substantially reduces the number of RIS configurations needed to achieve accurate localization of the target, a key step toward real-time imaging. In our current experimental setup, we can record signals from 128 configurations in approximately 1~s, corresponding to an imaging rate of $\sim$1~Hz (see Supplemental Video). The present limiting factor is the acquisition time of the vector network analyzer (VNA) and the data transfer rate between the VNA and the computer. Since the RIS itself can be reconfigured at speeds up to 1~kHz, true real-time imaging should be achievable with a dedicated hardware and optimized acquisition protocols. More information about the acquisition and computing times for the different experiments are given in the Materials and Methods.

\section*{Discussion}
This work introduces a novel method for estimating a reflection matrix with a single antenna and one or several programmable RIS. Our approach first introduces a characterization technique of the RIS via an optimization algorithm. Then, the sought-for reflection matrix can be obtained by pseudo-inversion. We have shown that this reflection matrix has the same features as the ones usually obtained in optics and acoustics with multi-elements arrays, thus, we effectively synthesize an array of antennas. {It confers an unexpected yet powerful advantage: the RIS gains a panoramic view of the entire scene, allowing non-line-of-sight imaging from the vantage point of a single emitter.}

Furthermore, we have been able to correctly predict the measured field received by the antenna for ``unmeasured" RIS configurations, i.e. configurations that were not used in the optimization process. We have shown that the estimated reflection matrices display all the expected properties. It is worth noting that the presented technique is robust despite the crude and approximate physical model of point-like scatterers utilized herein. More precisely, in \figeq{Eq.~(\ref{eq: psi})} a lot of paths are neglected. We assume that the antenna does not directly illuminate the scene (NLOS geometry) and therefore the contribution of triangular paths that interact once with the RIS and once with the scene (i.e. antenna $\rightarrow$ RIS $\rightarrow$ scene $\rightarrow$ antenna and the reciprocal) are negligible. We also assume that there is no multiple scattering between the RIS and the scene or the antenna. In our case, these assumptions are warranted because the waves propagate in free-space, but in complex environments a more complete description of the system will be required.

This opens up exciting perspectives for matrix-based techniques with any type of wave in regimes in which large arrays are unavailable but programmable metasurfaces such as RIS are accessible. For example in the microwave regime, wireless communication between devices can be improved without requiring feedback from them. This is possible as long as the devices act as guidestars (i.e. very bright point-like scatterers) within the medium, enabling the retrieval of the focusing law on each of them. In addition, guidestar-free methods are emerging, for example via a local perturbation such as movement~\cite{ambichlFocusingDisorderedMedia2017} or non-linearity~\cite{goicoechea2025detecting}, therefore strongly reflecting devices may not even be necessary. From an imaging standpoint, being able to estimate a reflection matrix also lets one take advantage of techniques inspired by adaptive optics to compensate for distortions~\cite{Badon2020, lambertDistortionMatrixApproach2020_2, yoonDeepOpticalImaging2020}.

Looking ahead, improvements can be made to accelerate the acquisition towards true real-time imaging. The same method can also be applied at different frequencies in order to obtain a broadband reflection matrix, thus making it possible to achieve axial resolution, as well as using  powerful aberration correction algorithms. Finally, using the same formalism, the full vector character of electromagnetic or elastic waves can be taken into account instead of the scalar approximation, thus unlocking additional degrees of freedom by controlling the polarization.

\section*{Materials and Methods}
\subsubsection*{Experimental Design}
The horn antenna used in the experiments has an aperture of $8.5 \times 13~\mathrm{cm^2}$. The experiments are performed at $f = 4.1$ GHz (wavelength $\lambda = 7.3$ cm). The antenna is located about $7\lambda \simeq 50$ cm away from the RIS and the target (when present) is located approximately at the same distance. During the calibration step, only the antenna and RIS are present in the environment; for the imaging step the targets are placed on top of polystyrene (transparent at that frequency) or fixed on a robot arm for the moving target experiment.

\subsubsection*{Reconfigurable intelligent surface}
The two RIS are prototypes purchased from Greenerwave consisting of $8\times 8$ thin resonant elements separated by approximately half a wavelength. Each element is controlled independently with two states available (polarizabilities $\alpha_0$ or $\alpha_1$).

\subsubsection*{Data Processing}
All processing of the experimental data were performed using Python custom written codes. The PyTorch framework~\cite{anselPyTorch2Faster2024} was used for the optimization process required in the calibration step. For an optimization with $N=2^{14}$ configurations and 8256 variables to estimate (two RIS), the algorithm takes approximately 45 minutes to run on CPU (Intel\textregistered\, Xeon\textregistered\, Gold 6230R). In the case of $N=1.5\times10^4$ and 2080 variables to estimate (one RIS), it takes approximately 5 minutes. For the retrieval of the reflection matrix, the pseudo-inverse takes about 3.5 minutes with $N=2.5\times10^4$ and a $128\times 128$ matrix to compute, 8 seconds with $N=1.5\times10^4$ and a $64\times 64$ matrix. In the real-time experiment for a single target the pseudo-inversion takes 80 milliseconds with $N=128$.


\bibliography{MatIm_RIS}

\bibliographystyle{Science}



\section*{Acknowledgements}
\begin{figure}[H]
    \centering
    \includegraphics[width=.8\textwidth]{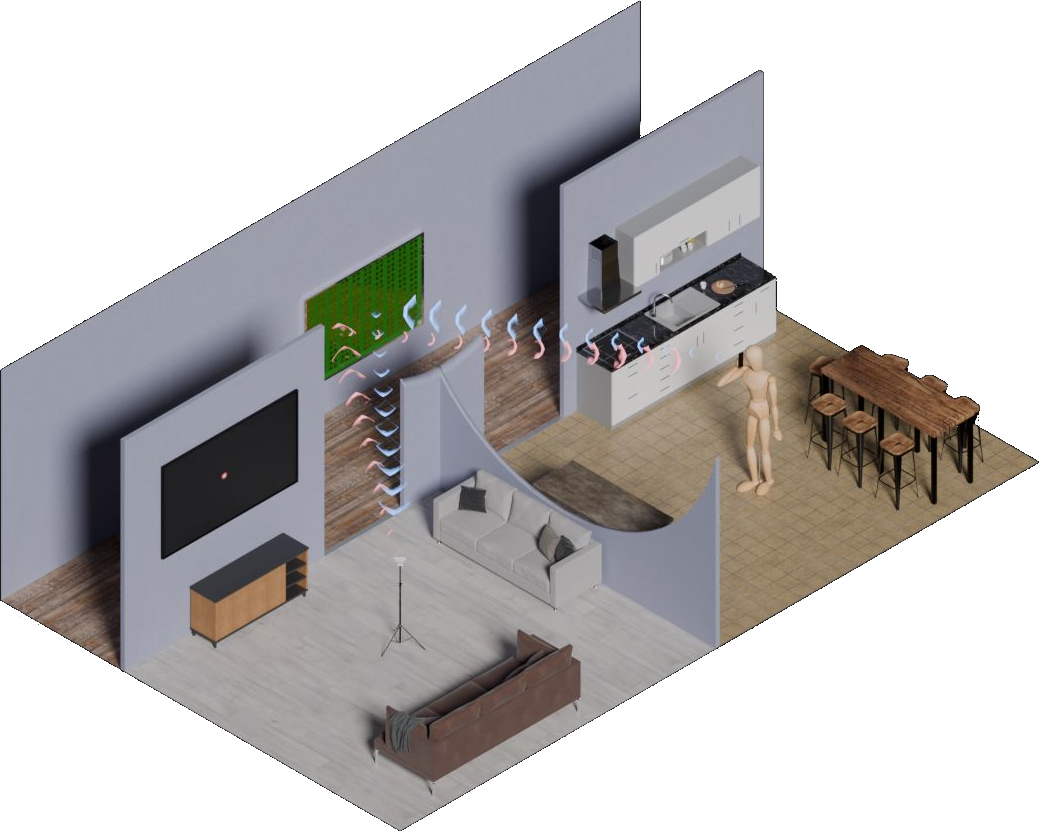}
    \caption{\label{fig:principle} {\bf Principle of the method.} Schematic of the single-antenna system coupled to a RIS for retrieving the reflection matrix in a non-line-of-sight configuration. The antenna’s emitted signal is scattered by the RIS to illuminate the scene, and the back-scattered field—reflected again by the metasurface—is recorded for multiple RIS configurations. From these measurements, the reflection matrix in the RIS-element basis is reconstructed, enabling the computation of the scene image.}
\end{figure}

\paragraph*{Funding:} {This work is supported in part by the European Union through a European Regional Development Fund (ERDF), by the Ministry of Higher Education and Research, CNRS, Brittany region, Conseils Départementaux d’Ille-et-Vilaine and Côtes d’Armor, Rennes Métropole, and Lannion Trégor Communauté, through the CPER Project CyMoCod, in part by the French ``Agence Nationale de la Recherche" (ANR) under Grant ANR-22-CPJ1-0070-01 and ANR-24-CE91-0007-01 for the project META-INCOME and in part by the Direction Générale de l'Armement through the Creach Labs under the project TRsweep. M.D. acknowledges the Institut Universitaire de France. M.F. and F.L. have received support from the Simons Foundation/Collaboration on Symmetry-Driven Extreme Wave Phenomena. The metasurface prototypes were purchased from Greenerwave. The authors acknowledge P. E. Davy for the 3D rendering of the concept in Fig.~\ref{fig:principle}.}

\paragraph*{Author contributions:} {M.F., F.L. and M.D. conceived and initiated the project. A.G. and M.D. performed and analyzed the experiments the experiments. A.G. did the theoretical modeling. A.G. and M.D. prepared the first draft of the manuscript. All authors discussed the results and contributed to finalizing the manuscript.}

\paragraph*{Competing interests:} {M.F. is one of the founders of the company, Greenerwave, that provided the RIS. M.F and F.L. are both scientific consultants of the same company.}
\paragraph*{Data availability:} All data needed to evaluate the conclusions in the paper are present in the paper and/or the Supplementary Materials. Additional data related to this paper may be requested from the authors.


\clearpage

\clearpage
\begin{figure}[H]
    \centering
    \includegraphics[width=\textwidth]{figures/figure_1.pdf}
    \caption{\label{fig:setup} {\bf Experimental setup.} (\textbf{A}) Photograph of the experiment for the matrix imaging of a single metal target with one $8\times 8$ RIS and a horn antenna. (\textbf{B}) Schematic illustration of matrix imaging with a single antenna and a RIS. We consider a microwave horn antenna illuminating a RIS (B1), the wave then bounces off the RIS toward the scattering scene to image (B2). The wave then travels back to the RIS (B3) before being detected with the same antenna (B4). For the calibration measurement, we remove the scattering medium so that the wave follows only (B1) and (B4).}
\end{figure}

\clearpage
\begin{figure}[H]
    \centering
    \includegraphics[width=\textwidth]{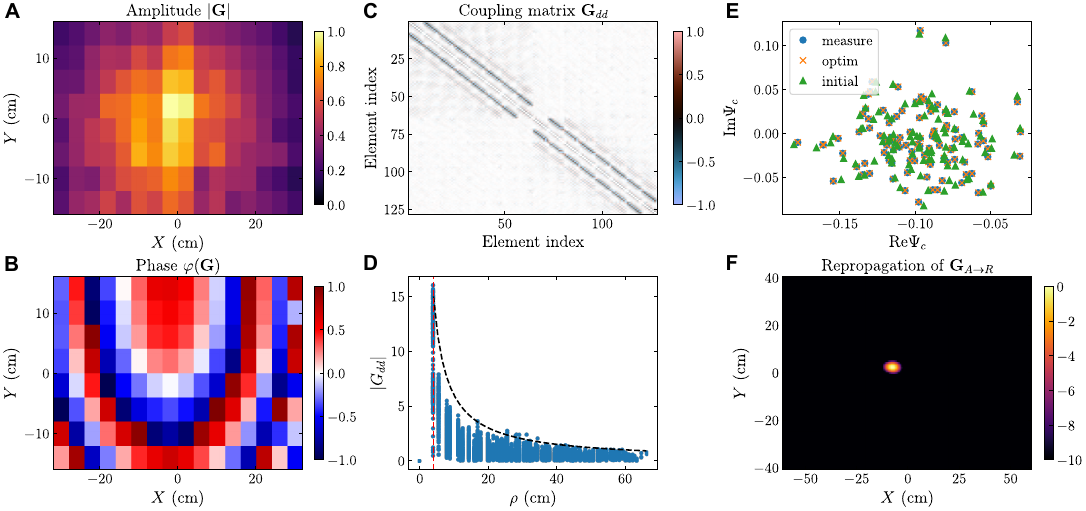}
    \caption{\label{fig:calibration} {\bf Calibration.} (\textbf{A} and \textbf{B}) Normalized amplitude and phase respectively of $\g{G}$. Despite being in a room, the wave propagates roughly as in free space. (\textbf{C}) Matrix of interaction between the elements of the RIS $\g{G}_{dd}$. The color represents the phase and the amplitude is coded in the transparency. (\textbf{D}) Amplitude of the coupling between elements as a function of their distance. The red dashed line represents the minimum distance $\rho$ between two elements; the black dashed line corresponds to $1/\rho$ which should be followed for pure radiative coupling. (\textbf{E}) Comparison of the estimated field with the experimental values; the mean absolute error is $5.4\%$ with the initial values and $0.2\%$ after optimization. (\textbf{F}) Image obtained by back-propagating the vector $\g{G}$, which corresponds to the location of the antenna.}
\end{figure}

\clearpage
\begin{figure}[H]
    \centering
    \includegraphics[width=\textwidth]{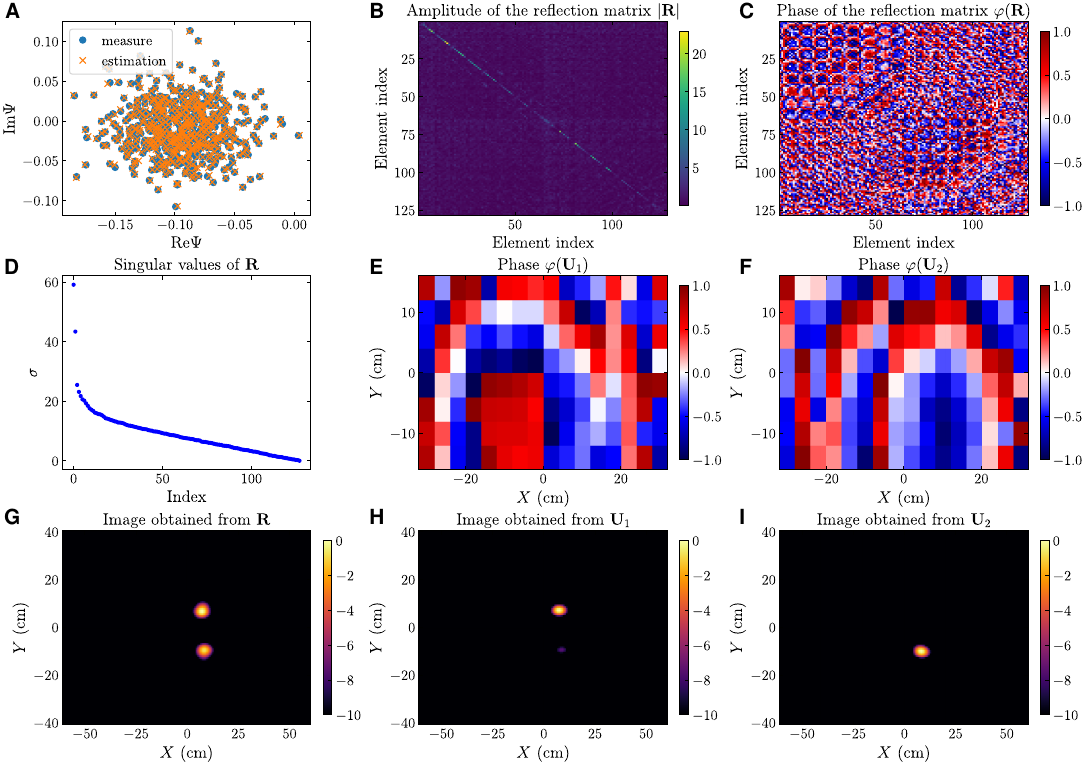}
    \caption{\label{fig:2targets} {\bf Imaging two targets.} (\textbf{A}) Comparison of the estimated field with the experimental values; the mean absolute error is $1.2\%$. (\textbf{B} and \textbf{C}) Respectively, amplitude and phase of the estimated reflection matrix $\g{R}$. (\textbf{D}) Plot of the singular values $\sigma$ of $\g{R}$. (\textbf{E} and \textbf{F}) Phase of the first two singular vectors $\g{U}_1$ and $\g{U}_2$ respectively, associated with the largest singular values. (\textbf{G}) Confocal image obtained from $\g{R}$. (\textbf{H} and \textbf{I}) Images obtained from the first two singular vectors $\g{U}_1$ and $\g{U}_2$ respectively.}
\end{figure}

\clearpage
\begin{figure}[H]
    \centering
    \includegraphics[width=\textwidth]{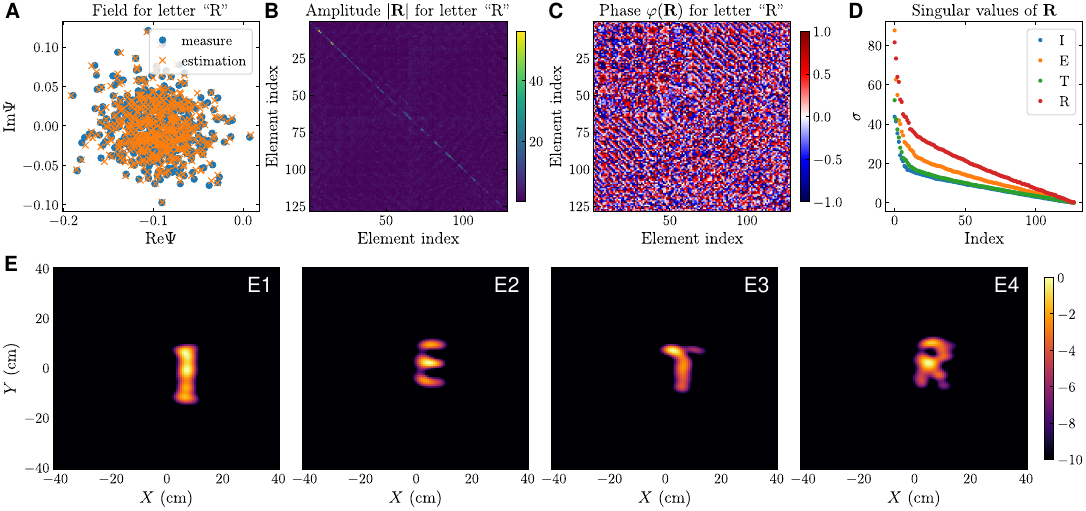}
    \caption{\label{fig:IETR} {\bf Imaging complex targets.} (\textbf{A}) Comparison of the estimated field with the experimental values for the most complex target in the shape of the letter ``R". The mean absolute errors for each of the letters of ``IETR" are $1.4\%$, $2.1\%$, $1.3\%$ and $3.3\%$ respectively. (\textbf{B} and \textbf{C}) Respectively, amplitude and phase of the estimated reflection matrix $\g{R}$ for the target ``R". (\textbf{D}) Singular values $\sigma$ of $\g{R}$ for each target. (\textbf{E}) Images obtained from $\g{R}$ for each target: ``I" (E1), ``E" (E2), ``T" (E3) and ``R" (E4).}
\end{figure}

\clearpage
\begin{figure}[H]
    \centering
    \includegraphics[width=\textwidth]{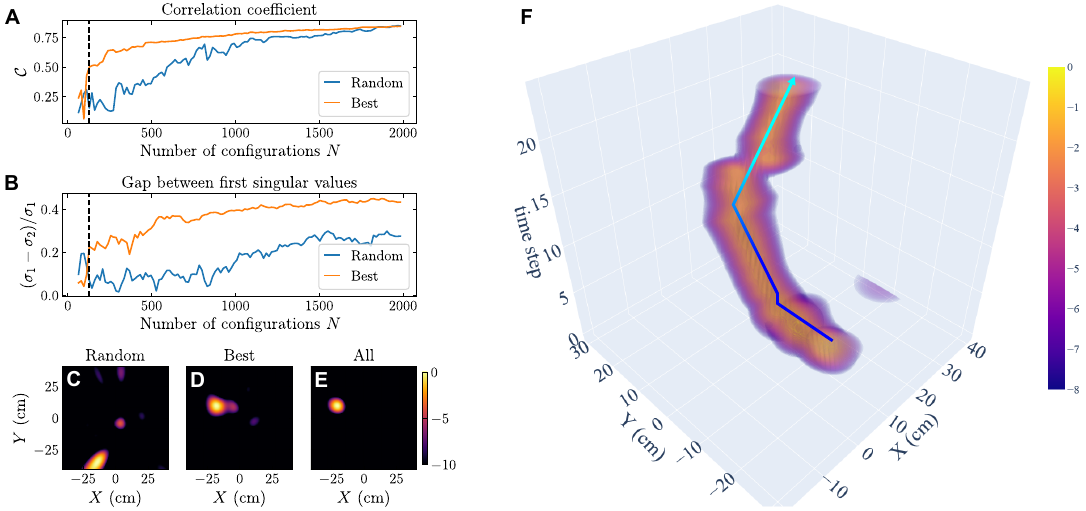}
    \caption{\label{fig:Moving} {\bf Moving target.} (\textbf{A}) Correlation coefficient $\mathcal{C} = |\g{U}_{1} \cdot \g{U}_{1,\mathrm{ref}}|$, where $\g{U}_1$ is estimated from $N$ configurations and $\g{U}_{1,\mathrm{ref}}$ is a reference obtained from all configurations. We estimate $\g{R}$ by choosing configurations randomly or with a pre-selection (see main text). (\textbf{B}) Gap between the first two singular values $\sigma_1 - \sigma_2$ for each case. (\textbf{C} and \textbf{D}) Images obtained for {$N=128=2N_d$} configurations by randomly choosing the configurations or by pre-selecting them, respectively. (\textbf{E}) Image obtained by taking into account all of the configurations. (\textbf{F}) 3D plot of the trajectory obtained by estimating $\g{U}_1$ from $N=1000$ configurations. The images obtained form a trajectory that follows the real path of the target represented by the blue line (early times in dark blue, late times in cyan).}
\end{figure}

\clearpage
\begin{center}
{\Large Supplementary Materials:}\\
{\LARGE Single-Antenna Non-Line-of-Sight Matrix Imaging via  Reconfigurable Intelligent Surfaces}\\[\baselineskip]

\normalsize{Antton Goïcoechea$^{1\dagger\ast}$, François Sarrazin$^1$, Theodosios Karamanos$^{2,3}$, \\ Mathias Fink$^4$, Fabrice Lemoult$^4$, Matthieu Davy$1^{\ast}$\\[\baselineskip]

\normalsize{$^{1}$Université de Rennes, CNRS, IETR, Rennes, France}\\
\normalsize{$^{2}$Sorbonne Université, CNRS, Laboratoire GeePs, Paris, France}\\
\normalsize{$^{3}$Université Paris-Saclay, CentraleSupélec, CNRS, Laboratoire GeePs, Gif-sur-Yvette, France}\\
\normalsize{$^{4}$Institut Langevin, ESPCI Paris, Université PSL, CNRS, Paris, France}\\[\baselineskip]

\normalsize{$\dagger$Present address: Institut Jean Le Rond d'Alembert, Sorbonne Université, CNRS, Paris, France}\\
\normalsize{$\ast$Corresponding authors. E-mail: antton.goicoechea@dalembert.upmc.fr (A.G.)};\\ \ \normalsize{matthieu.davy@univ-rennes.fr (M.D.)}
}
\end{center}

\section*{Supplementary Materials}
\textbf{The PDF file includes:}
\begin{itemize}
    \item[] Supplementary text
    \item[] References
    \item[] Figs. S1 to S2
\end{itemize}


\section{Optimization in the calibration step}
\subsection{Optimization algorithm}
The retrieval of the vectors and matrices is performed using the Adam optimization scheme~\cite{kingmaAdamMethodStochastic2014} available in Pytorch. We use $N = 2^{14} = 16384$ random RIS configurations so that $N > N_p$, where $N_p$ is the number of variables we are trying to optimize (which depends on $N_d = 128$ in our experiments). Note that this number is somewhat arbitrary, but it allows for decent calculation speed while making sure we have enough `measurements' of the field.

In the calibration geometry we are computing:
\begin{equation}\label{eqS: field calib opt}
    \hat{\Psi}_c = \hat{\Psi}_{0} + \hat{\g{G}} \frac{\hat{\boldsymbol{\alpha}}}{\mathbf{1} - \hat{\g{G}}_{dd}\hat{\boldsymbol{\alpha}}} \hat{\g{G}}^T
\end{equation}
where $\hat{\g{x}}$ corresponds to the estimate of $\g{x}$. The loss function we use is the mean absolute error:
\begin{equation}\label{eqS: loss calib}
    \mathcal{L}(\Psi_c,\hat{\Psi}_c) = \left\langle \left| \Psi_c - \hat{\Psi}_c \right| \right\rangle_b
\end{equation}
where $\left\langle\cdots\right\rangle_b$ represents average over batch configurations. Since $\hat{\g{G}}$, $\hat{\boldsymbol{\alpha}}$ and $\hat{\g{G}}_{dd}$ can be scaled by an arbitrary complex factor, we choose to fix $\hat{\alpha}_0 = 1/k_0$ and write $\hat{\alpha}_1 = \alpha_0 \hat{r} e^{i\hat{\theta}}$. Because of the symmetry of $\g{G}_{dd}$, we compute only its upper triangular part, thus reducing the number of (real) variables to optimize to $4 + N_d + N_d^2 = 16516$ for $N_d = 128$.

\subsection{Initialization}
To initialize the problem with a decent guess we develop Eq.~\eqref{eqS: field calib opt} up to the second order:
\begin{equation}\label{eqS: field calib init}
    \hat{\Psi}_c = \hat{\Psi}_{0} + \hat{\g{G}} \hat{\boldsymbol{\alpha}} \hat{\g{G}}^T + \hat{\g{G}} \hat{\boldsymbol{\alpha}} \hat{\g{G}}_{dd}\hat{\boldsymbol{\alpha}} \hat{\g{G}}^T
\end{equation}
We then set $\hat{\alpha}_1^0 = -\hat{\alpha}_0$ and $\hat{\Psi}_{0}^0 = \langle \Psi_c \rangle$. $\hat{\g{G}}^0$ is obtained by solving Eq.~\eqref{eqS: field calib init} with the first two terms only. After this step, we get $\hat{\g{G}}_{dd}^0$ by solving Eq.~\eqref{eqS: field calib init}. For the latter case we use the following property:
\begin{align}
    y
    &= \g{M} \g{X} \g{M}^T \nonumber \\
    &= \Tr \left[ \g{M}^{T} \g{X} \g{M} \right] \nonumber \\
    &= \mathrm{vec}\left( \g{M}^{T} \g{M} \right) \cdot \mathrm{vec}\left( \g{X}^{*} \right)
\end{align}
By concatenating the vectors $\mathrm{vec}\left( \g{M}^{T} \g{M} \right)$ in a matrix $\g{B}$ for $N$ realizations, the vector $\g{y}$ containing the $N$ realizations of $y$ is written:
\begin{equation}
    \g{y} = \hat{\g{B}} \,\mathrm{vec}\left( \g{X} \right)
\end{equation}
$\g{X}$ can then be obtained with a least-square solver.

\subsection{Phase correction}
Because $\hat{\g{G}}$ appears twice in Eq.~\eqref{eqS: field calib opt}, each component of the vector is defined up to a $\pi$ phase shift and this phase shift is thus also in $\hat{\g{G}}_{dd}$. We assume that the RIS should be in the far field of the antenna for the phase across the RIS to be slowly varying. This is a weak limitation since the antenna should illuminate the whole surface anyway. We correct the phase by minimizing the phase difference $\Delta\phi_n$ between neighboring elements of the RIS in $\hat{\g{G}}$ as well as its first and second derivatives, $\Delta\phi_n'$ and $\Delta\phi_n''$ respectively, for smoothness. More precisely, we need to find the vector of binary weights $\g{p}$ that minimize the loss function
\begin{equation}
    \mathcal{L}(\g{p}) = \frac{1}{N_d-1}\sum_{n=1}^{N_d-1} \left\{ \sin^2\left[\frac{\Delta\phi_n(\g{p})}{2}\right] + \sin^2\left[\frac{\Delta\phi_n'(\g{p})}{2}\right] + \sin^2\left[\frac{\Delta\phi_n''(\g{p})}{2}\right] \right\}
\end{equation}
where $\Delta\phi_n(\g{p}) = \phi_{n+1}(p_{+1})- \phi_n(p_n)$ with $\phi_n(p_n) = \phi_n + p_n \pi$.

The total number of combinations is $2^{N_d-1}$, so we cannot try all possible $\g{p}$. Therefore, we first compute the loss function $\mathcal{L}(\g{p})$ for each rows of $\hat{\g{G}}$ and then a column.


\section{Reflection matrix}
\subsection{Field estimation by pseudo-inversion}
We are now looking to estimate the reflection matrix $\hat{\g{R}}$ using
\begin{equation}\label{eqS: psi imag opt}
    \hat{\Psi}_{\mathcal{I}} = \Psi - \Psi_c = \hat{\g{G}} \, \hat{\boldsymbol{\Gamma}} \, \hat{\g{R}} \, \hat{\boldsymbol{\Gamma}} \, \hat{\g{G}}^T
\end{equation}
Again, because of the symmetry of $\g{R}$, we compute only its upper triangular part (including its diagonal), so that the total number of (real) variables to estimate is $N_d (N_d +1) = 16512$ for $N_d = 128$.

We can write the measured fields (with the calibration subtracted) for each configuration as
\begin{align}
    \Psi_{\mathcal{I}}
    &= \g{G} \, \boldsymbol{\Gamma} \, \g{R} \, \boldsymbol{\Gamma} \, \g{G}^T \nonumber \\
    &= \g{A} \g{R} \g{A}^{T} \nonumber \\
    &= \Tr \left[ \g{A}^{T} \g{A} \g{R} \right] \nonumber \\
    &= \mathrm{vec}\left( \g{A}^{T} \g{A} \right) \cdot \mathrm{vec}\left( \g{R}^{*} \right)
\end{align}
with $\g{A} = \g{G} \boldsymbol{\Gamma}_R$.

The field can then be written as a vector for all RIS configurations in the form:
\begin{equation} \label{eqS: field B}
    \boldsymbol{\Psi}_{\mathcal{I}} = \g{B}\,\mathrm{vec}\left( \g{R} \right)
\end{equation}
where $\g{B}$ is a $N\times (N_p/2)$ matrix built by concatenating the vectors $\mathrm{vec}\left( \g{A}^{T} \g{A} \right)$ for $N$ configurations. The reflection matrix $\hat{\g{R}}$ is then simply obtained by pseudo-inversion of $\hat{\g{B}}$:
\begin{equation}
    \mathrm{vec}\left( \hat{\g{R}} \right) = \hat{\g{B}}^{+} \boldsymbol{\Psi}_{\mathcal{I}}
\end{equation}
We emphasize here that as long as the calibration remains the same, one can compute a reflection matrix $\hat{\g{R}}$ by just making a matrix product between the measurements and $\hat{\g{B}}^{+}$, which is a very fast operation numerically.

\subsection{Regularization of the pseudo-inverse}
The pseudo-inverse is computed using the singular value decomposition, therefore, some cutoff $\xi$ should be used in order not to give too much weight to small singular values which should ideally be zero. The singular values of $\g{B}$ are shown in Fig.~\figeq{S1A}, and we see that the drop happens around $\sigma \sim 10^{-3}$. In Fig.~\figeq{S1} we show the results for different values of the absolute tolerance $\xi$, below which the singular values are considered to be zero before the pseudo-inversion, when computing the reflection matrix associated to the ``R" shaped target. As expected, if $\xi$ is too low, the resulting reflection matrix $\g{R}$ is nonphysical and the image meaningless. On the other hand, if $\xi$ is too large, not enough singular values are taken into account in the pseudo-inversion and some information is lost. For the figures presented in the main text and in the following section, we use an absolute tolerance of $\xi = 2\times10^{-3}$.

\subsection{Number of configurations taken into account}
We now check the precision of our estimation of the field with the ``R" shaped target, which is the most complex one, as a function of the number of RIS configurations $N$ taken into account. The mean absolute error between measurement $\boldsymbol{\Psi}$ and estimation $\hat{\boldsymbol{\Psi}}$ is given by
\begin{equation}
\mathrm{MAE} = \left\langle \left| \frac{\boldsymbol{\Psi} - \hat{\boldsymbol{\Psi}}}{\boldsymbol{\Psi}} \right| \right\rangle
\end{equation}
The results for various $N$ are shown in Fig.~\figeq{S2}, in which we clearly see that for such a complex target, a high number of illuminations are required to obtain $\g{R}$. We also readily see that taking $N$ low allows only for the reconstruction of the brightest part of the target. This allows us to make images quickly for low-complexity targets, as discussed in the main text.

\section{Triangular paths}
In Eq.~1 of the main text, triangular paths that interact once with the RIS and once with the scene (i.e. antenna $\rightarrow$ RIS $\rightarrow$ scene $\rightarrow$ antenna and the reciprocal) are neglected because we assume that the antenna does not directly illuminate the scene. In general the detected field is
\begin{equation}
    \Psi = \Psi_c + \Psi_{\mathcal{I}} + \Psi_{\bigtriangleup}
\end{equation}
where
\begin{equation}
     \Psi_{\bigtriangleup} = \g{G}_{A\rightarrow R} \, \boldsymbol{\Gamma} \, \g{G}_{R\rightarrow S} \, \boldsymbol{\gamma} \, \g{G}_{A\rightarrow S}^T
    + \g{G}_{A\rightarrow S} \, \boldsymbol{\gamma} \, \g{G}_{R\rightarrow S}^T \, \boldsymbol{\Gamma} \, \g{G}_{A\rightarrow R}^T
\end{equation}
For clarity we put indices below the Green's function to indicate between which component the waves propagates: $\g{G}_{i\rightarrow j}$ is the Green's function from $i$ to $j$, where $i$, $j$ can be the antenna ($A$), the RIS ($R$) or the scatterers of the scene ($S$).

Using $\g{R} = \g{G}_{R\rightarrow S} \, \boldsymbol{\gamma} \, \g{G}_{R\rightarrow S}^T$ and $\g{G}_{R\rightarrow S}^{\dagger} \, \g{G}_{R\rightarrow S}\simeq 1$ and reciprocity the previous equation can be rewritten as
\begin{equation}
     \Psi_{\bigtriangleup} = 2\times \g{G}_{A\rightarrow R} \boldsymbol{\Gamma} \, \g{R} \, \g{G}_{R\rightarrow S}^{*} \, \g{G}_{A\rightarrow S}^T
\end{equation}
In order for the procedure to work, we need $\Psi_{\bigtriangleup} \ll \Psi_{\mathcal{I}}$, i.e.
\begin{equation}
    \left\| \g{G}_{A\rightarrow S} \g{G}_{R\rightarrow S}^{\dagger} \right\| \ll \left\| \g{G}_{A\rightarrow R} \boldsymbol{\Gamma} \right\|
\end{equation}


\clearpage
\begin{figure}[H]
    \centering
    \includegraphics[width=\textwidth]{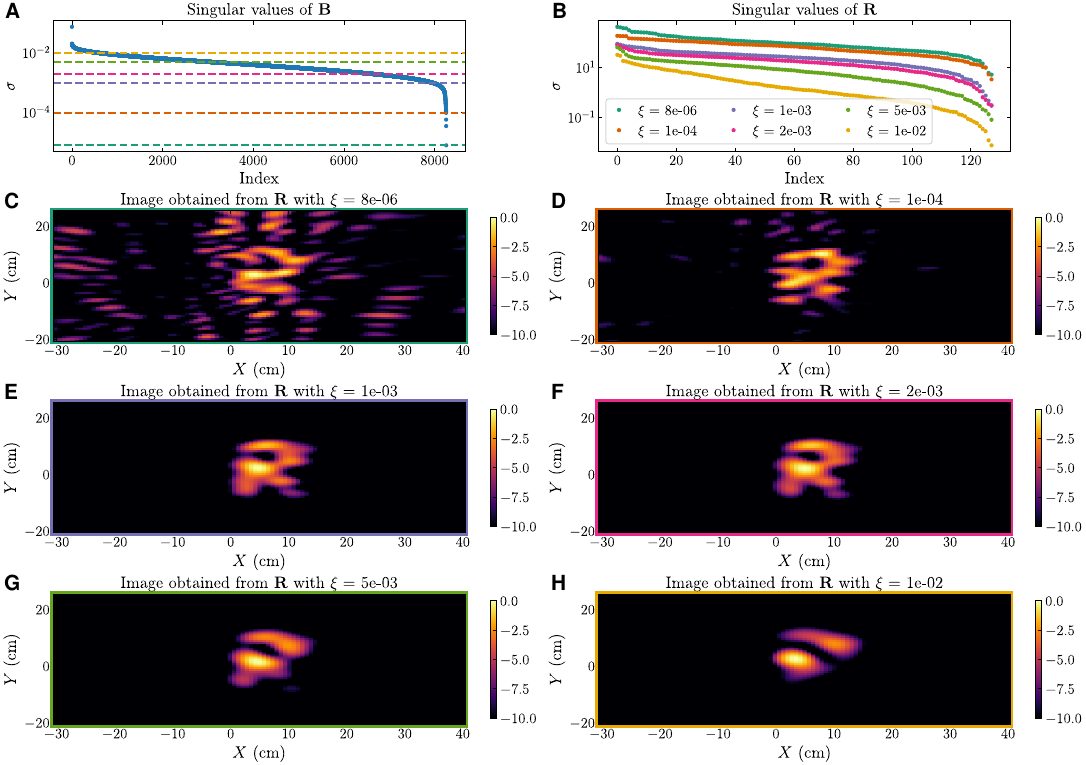}
\end{figure}

\noindent {\bf Fig. S1. Influence of the regularization process for the image reconstruction.} (\textbf{A}) SVD of the matrix $\g{B}$ appearing in Eq.~\eqref{eqS: field B}. The colored dashed lines corresponds to the threshold value $\xi$ below which the singular values are considered to be zero in the pseudo-inversion. (\textbf{B}) Singular values of $\g{R}$ computed with different values of $\xi$. (\textbf{C} - \textbf{H}) Reconstructed image from $\g{R}$ for different $\xi$.

\clearpage
\begin{figure}[H]
    \centering
    \includegraphics[width=\textwidth]{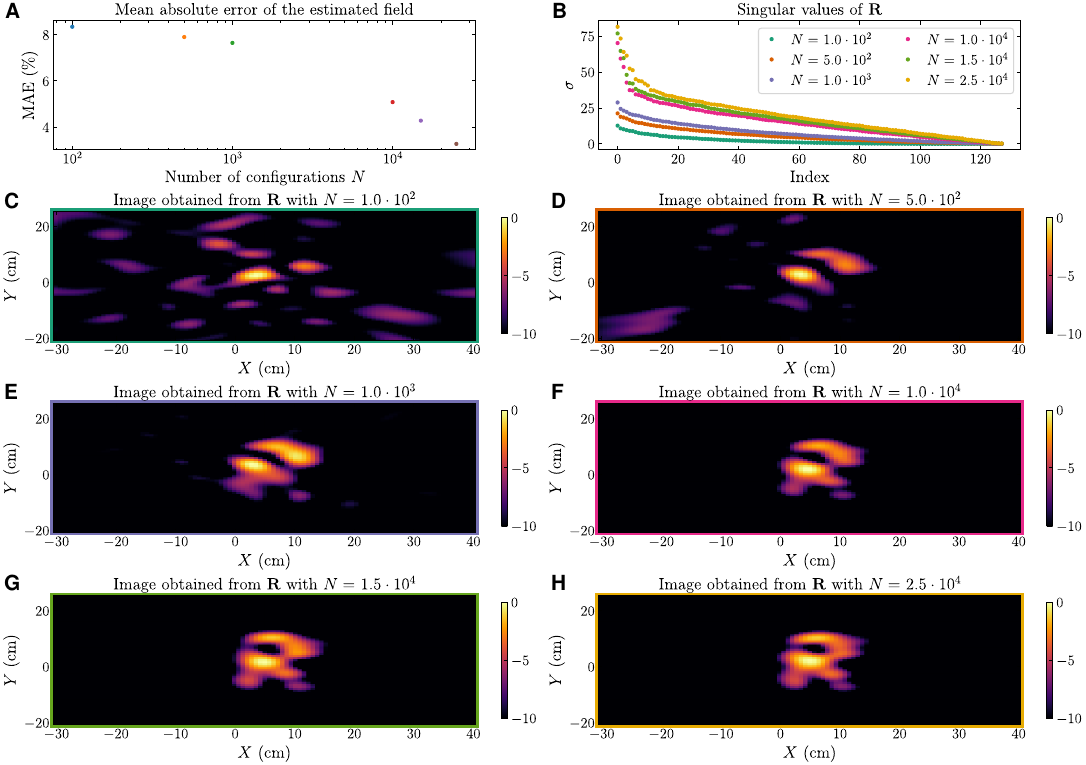}
\end{figure}

\noindent {\bf Fig. S2. Influence of the number of random configurations for the image reconstruction.} (\textbf{A}) Mean absolute error of the estimated field $\mathrm{MAE}$ as a function of the number of RIS configurations taken into account to compute $\mathbf{R}$. (\textbf{B}) Singular values of $\g{R}$ computed with different values of $N$. (\textbf{C} - \textbf{H}) Reconstructed image from $\g{R}$ for different $N$.

\end{document}